\documentclass[showkeys,nofootinbib,prd]{revtex4}

\usepackage{amsfonts}
\usepackage{amssymb}
\usepackage{natbib}
\usepackage[latin1]{inputenc}
\usepackage[babel]{csquotes}
\usepackage{graphicx}
\usepackage[english]{babel}

\usepackage[T1]{fontenc}
\usepackage[centertags]{amsmath}

\usepackage{mathrsfs}
\usepackage{amsmath}
\usepackage{amsthm}
\usepackage{amssymb}
\usepackage{mathtools}
\usepackage{wasysym}
\usepackage{subfigure}
\DeclareFontFamily{U}{mathb}{\hyphenchar\font45}
\DeclareFontShape{U}{mathb}{m}{n}{
      <5> <6> <7> <8> <9> <10> gen * mathb
      <10.95> mathb10 <12> <14.4> <17.28> <20.74> <24.88> mathb12
      }{}
\DeclareSymbolFont{mathb}{U}{mathb}{m}{n}

\DeclareMathSymbol{\Sun}{3}{mathb}{"40}
\newcommand\T{\rule{0pt}{4ex}}       
\newcommand\B{\rule[-4ex]{0pt}{0pt}} 

\allowdisplaybreaks

\begin{document}
\title{Gravitational waves from a binary source in higher dimensional spacetime with compactified extra dimensions}
\author{Mattia Villani}
\affiliation{University of Urbino Carlo Bo, Department of Pure and Applied Sciences (DiSPeA), Via Santa Chiara, 27, Urbino (PU), 61029, Italy}
\email{mattia.villani@uniurb.it}
\begin{abstract}
We consider the emission of gravitational waves (GW) from a compact binary in a spacetime with compactified extra dimensions. We solve the homogeneous and non-homogeneous wave equation, proving that there is an infinite sum of exponentially decaying pseudo-massive modes which add to the usual one which behaves as $1/r$ at infinity. We calculate the metric potentials, the equations of motion and the energy flux. We find that in the equations of motion there is a new term at -1PN order. This implies a modification of the energy flux. 
\end{abstract}

\maketitle

\section{Introduction}
In this paper, we shall tackle the Post Newtonian (PN) expansion of the metric and of the energy flux of {gravitational waves (GW)} in a $d+4$-dimensional spacetime, with generic $d$ for a binary system of compact objects. A similar work was carried out in \cite{multi}, however, here we consider the particular situation in which the $d$ extra dimensions are compactified to a hypersphere. The idea of the presence of extra dimensions comes from String Theory (see, for example, the introductory books \cite{st1,st2}). These dimensions must be compactified to a small volume in order to make contact with the everyday experience that macroscopically there are only 3 spatial dimensions plus time. The aim of the present work is to arrive to a waveform that could be used in data analysis in order to put limits on the number of extra dimensions and their size. For similar ideas, see, for example, \cite{ap1,ap2,ap3,ap4}. In this paper, we shall only consider the 1PN expansion; higher order expansions will be calculated in forthcoming papers.

We use the multipolar post-Minkowskian expansion developed during the years by Blanchet and collaborators; see \cite{LRR1} for a review, the more recent version \cite{LRR2} and references therein. In this formalism, the metric and energy flux were calculated up to 4PN order including spin effects (such as spin-spin and spin-orbit interactions). We also calculated the effects of the magnetic field in binary neutron stars and neutron star-black hole systems {using the post-Minkowskian expansion}  \cite{mio,mio2}.

The main result of this work is that we show that there is an infinite sum of exponentially decaying pseudo massive modes in addition to the usual one behaving as $1/r$ at spatial infinity and that in the equation of motion in the center of mass and for circular orbits a term appears at -1PN order; this has as a consequence that the energy flux at infinity is also modified. These additional terms depend on the size of the extra dimensions and on their number; therefore, by comparing the theoretical expectation with observations, one could set limits on these quantities.

This paper is organized as follows: after presenting our notation and the Einstein's equations in the next Section, in Section \ref{sec:homogeneous} we present the solution of the $d+4$-dimensional wave equation without a source, while in section \ref{sec:non-homog} we present the multipolar expansion of the perturbation tensor. In Section \ref{sec:green}, we calculate the Green's function of the multidimensional Laplacian. In Section from \ref{sec:met} to \ref{sec:flux} we calculate the 1PN expansion of the metric, the equations of motion, and the energy flux. {In Section \ref{sec:comp} we compare our results to previous works.} Finally, in Section \ref{sec:concl}, we conclude our presentation.

\section{Notation and Einstein's equations}

We consider a d+4-dimensional spacetime, i.e. a spacetime with d extra dimensions. We indicate the 4-dimensional indices with greek letters from the beginning of the alphabet $\alpha,\beta,\dots$, while we use capital latin letters $A,B,\dots$ for the extra dimensions. When we do not want to distinguish the indices we use greek letters from the middle of the alphabet $\mu,\nu,\dots$. We assume that the metric has the form:
\begin{equation}
    g_{\mu\nu}=\left( \begin{array}{cc}
    g_{\alpha\beta} & 0 \\
    0 & g_{AB}
    \end{array}\right)
\end{equation}
i.e., we assume that the metric is block diagonal without mixing between the indices. The flat metric is:
\begin{equation}
    \eta_{\mu\nu}=\left( \begin{array}{cc}
    \eta_{\alpha\beta} & 0 \\
    0 & \tilde{g}_{AB}
    \end{array}\right)
\end{equation}
where $\eta_{\alpha\beta}$ is the Minkowski metric, while $\tilde{g}_{AB}$ is the metric of a d-dimensional hypersphere.

Einstein's equations are written as usual:
\begin{equation}
    R_{\mu\nu}-\dfrac{R}{2}\,g_{\mu\nu}=\dfrac{8\pi G}{c^4}\,T_{\mu\nu},
\end{equation}
where $T_{\mu\nu}$ is the stress-energy tensor.

If we assume that the metric is perturbed by the presence of GW, we can write the perturbation as ({see equation (49) of \cite{LRR1}}):
\begin{equation}
    h_{\mu\nu}=\sqrt{-g}\,g_{\mu\nu}-\eta_{\mu\nu}.
\end{equation}
The Einstein's equations for the perturbation become:
\begin{equation}
    \Box h_{\mu\nu}=\dfrac{16\pi G}{c^4}\,\tau_{\mu\nu}, \qquad \tau_{\mu\nu}=|g|\,T_{\mu\nu}+\dfrac{c^4}{16\pi G}\,\Lambda_{\mu\nu}.
\end{equation}
The full expression of $\Lambda_{\mu\nu}$ {for higher dimensional spacetime is reported in equation (175) of} \cite{LRR1}. If we assume that matter is confined on the 4D spacetime, i.e. that $T_{AB}=0$, the Einstein's equations can be written as follows:
\begin{equation}
    \left\{ \begin{array}{l}
    \Box h_{\alpha\beta}=\dfrac{16\pi G}{c^4}\,|g|\,T_{\alpha\beta}+\Lambda_{\alpha\beta},\\
    \Box h_{AB}=\Lambda_{AB}
    \end{array}\right.
\end{equation}

{Finally, in the following we shall use the harmonic gauge, defined in equation (3.2) of \cite{art2}.}

\section{General solution of the past stationary linearized equations}
\label{sec:homogeneous}

We consider the homogeneous wave equation:
\begin{equation}\label{eq:homog}
    \Box h^{\mu\nu}=\left( \dfrac{1}{c^2}\,\partial_t^2-\Delta-\Delta_d \right)\,h^{\mu\nu}=0,
\end{equation}
where we have separated the dependence over the extra dimensions and we have {(see equation (2.3) in \cite{higuchi})}:
\begin{equation}
    \Delta_d=\dfrac{1}{\sqrt{\tilde{g}_d}}\partial_A\left( \sqrt{\tilde{g}_d}\,\partial^A \right)
\end{equation}

Following \cite{higuchi,higuchi2,rubin}, we have that:
\begin{equation}
    \Delta_d\,Y_{l_1,\dots,l_d}(\vec{\theta})=\dfrac{1}{\sqrt{\tilde{g_d}}}\partial_A\left( \sqrt{\tilde{g}_d}\,\partial^A Y_{l_1,\dots,l_d}(\vec{\theta}) \right)=-l_d(l_d+d-1)\,Y_{l_1,\dots,l_d}(\vec{\theta})
\end{equation}
where $Y_{l_1,\dots,l_d}(\vec{\theta})$ are the eigenfunctions of the d-dimensional Laplacian and $l_1,\dots,l_d$ are integers such that ${l_d>l_{d-1}>\dots>l_2>|l_1|}$. The expression of the eigenfunction is complicated but can be found in {equations (2.5)--(2.8) of \cite{higuchi}, but see also \cite{higuchi2,rubin}}. We have indicated the dependence over the $d$ angular variable $\theta_d$ as a vector $\vec{\theta}$. 

We now introduce a notation similar to \cite{thorne}. First we transform the $d$ angular variable to $d+1$ cartesian coordinates:
\begin{align}
x_1&=R\sin\theta_d\sin\theta_{d-1}\dots\sin\theta_2\sin\theta_1,\\
x_2&=R\sin\theta_d\sin\theta_{d-1}\dots\sin\theta_2\cos\theta_1,\\
x_3&=R\sin\theta_d\sin\theta_{d-1}\dots\cos\theta_2,\\
&\dots\\
x_{d} &=R\sin\theta_d\cos\theta_{d-1},\\
x_{d+1} &=R\cos\theta_d.
\end{align}
We also define the $d+1$ vectors
\begin{equation}\label{eq:N}
N_A=\dfrac{x_A}{R}.
\end{equation}

If we write $h^{\mu\nu}= \tilde{h}^{\mu\nu}_{\vec{l}}\,N_{\vec{l}}$, where $N_{\vec{l}}$ is a symmetric trace-free tensor and where we use the multi-index notation $\vec{l}{=A_1 A_2\dots A_{l_d}}$ {where each $A_i$ takes values in ${1,2,\dots (d+1)}$}, we obtain from equation \eqref{eq:homog}:
\begin{equation}\label{eq:homog2}
    \left( \dfrac{1}{c^2}\,\partial_t^2-\Delta+\dfrac{l_d(l_d+d-1)}{R^2} \right)\,\tilde{h}^{\mu\nu}_{\vec{l}}=0,
\end{equation}
where $R$ is the size of the extra dimensions. Thus, we have transformed the original d+4 dimensional wave equation into a 4 dimensional Klein-Gordon equation with pseudo-mass term $m_d=\sqrt{l_d(l_d+d-1)}/R$. The boundary and initial conditions are ({see equations (2.3)--(2.5) in \cite{art1}}):
\begin{align}
    &\partial_\mu \tilde{h}^{\mu\nu}_{\vec{l}}=0,\\
    &\partial_t\, \tilde{h}^{\mu\nu}_{\vec{l}}=0, \quad t\leq -T,\\
    &\lim_{r\rightarrow \infty}\tilde{h}^{\mu\nu}_{\vec{l}}=0, \quad t\leq -T.
\end{align}
Now we write $\tilde{h}^{\mu\nu}_{\vec{l}}= \tilde{h}^{\mu\nu}_{\vec{l}L}\,n_L$, where \cite{thorne}
\begin{equation}
    n_i=\dfrac{x_i}{r}
\end{equation}
{and where we have used the multi-index notation ${L=i_1i_2i_3\dots i_l}$ (see \cite{art1} for a similar notation}. We substitute it in \eqref{eq:homog2}, thus finding:
\begin{equation}
    \left( \dfrac{1}{c^2}\,\partial_t^2-\dfrac{1}{r^2}\partial_r\left(r^2\partial_r\right)+\dfrac{l(l+1)}{r^2}+m_d^2 \right)\,\tilde{h}^{\mu\nu}_{\vec{l},L}=0.
\end{equation}

We introduce the retarded and advanced coordinates
\begin{equation}
    u=t-r/c, \qquad v=t+r/c
\end{equation}
and consider $f^{\mu\nu}_{\vec{l}L}(u,v)=\tilde{h}^{\mu\nu}_{\vec{l}L}/(v-u)^l$, thus finding
\begin{equation}
    (v-u)\,\left(\partial_{uv}\,f^{\mu\nu} + m_d^2 f^{\mu\nu} \right)-(l+1)\,\partial_v\,f^{\mu\nu}+(l+1)\partial_u\,f^{\mu\nu}=0
\end{equation}
{The solution for generic $l$ is:}
\begin{equation}
    f^{\mu\nu}_{\vec{l}L}=A^{\mu\nu}\,(v-u)^{-1/2-l} J_{1/2+l}[i\,m_d\,(v-u)] +B^{\mu\nu}\, (v-u)^{-1/2-l} Y_{1/2+l}[i\,m_d\,(v-u)],
\end{equation}
{where $J$ and $Y$ are the Bessel functions. The Bessel functions can be expanded in terms of exponentials using for example Mathematica\footnote{One needs to fix $l$ to get the expansion. It seems Mathematica cannot handle generic $l$.}; the constants $A^{\mu\nu}$ and $B^{\mu\nu}$ must be chosen in such a way as to eliminate the growing exponential ${\exp(m_d\,(v-u))=\exp(2m_d\,r/c)}$: we find that $B^{\mu\nu}=-i\,A^{\mu\nu}$ suits this need. The expression of $f^{\mu\nu}_{\vec{l}L}$ for values of $l$ from 0 to 4 is given in table \ref{tab:expr}. Thus, the solution for $m_d\neq 0$ is an exponentially decaying function, while for $m_d=0$ is the usual inverse power of $(v-u)$. This means that $h^{\mu\nu}_L$ is an infinite sum of fast decaying function, except for the $m_d=0$ mode (corresponding to $l_d=0$) which radiates to infinity. It would be interesting to find out whether the fast decaying solutions leave any imprint on the wave form at infinity. The expression for $\tilde{h}^{\mu\nu}_{\vec{l}L}$ is:\footnote{It could also be written in terms of the Hankel functions.}}
\begin{subequations}
\begin{equation}
    \tilde{h}^{\mu\nu}_{\vec{l}L}=\Big[{A^{\mu\nu}\,(v-u)^{-1/2-l} J_{1/2+l}[i\,m_d\,(v-u)] -i\,A^{\mu\nu}\, (v-u)^{-1/2-l} Y_{1/2+l}[i\,m_d\,(v-u)]}\Big]\,{(v-u)^l}.
\end{equation}
\begin{equation}
    h_L=\sum_{l_d}\,\tilde{h}_{\vec{l}L}\,N_{\vec{l}}
\end{equation}
\end{subequations}
{where we use the notation $\sum_{l_d}$ to indicate a sum over all the eigenvalues relative to the extra dimensions.} {Finally, we note that the first two derivatives of the function $f^{\mu\nu}$ are given by:}
\begin{align}
    \partial_{u}f^{\mu\nu}_{\vec{l}L}&=\dfrac{i\,m_d\,f^{\mu\nu}_{\vec{l}L+1}}{v-u},\\
    \partial^2_{u}f^{\mu\nu}_{\vec{l}L}&=\dfrac{i\,m_d\,\left(-i\,f^{\mu\nu}_{\vec{l}L+1}\,(v-u)+f^{\mu\nu}_{\vec{l}L+2}\right)}{(v-u)^2}
\end{align}

\begin{table}[ht]
    \centering
    \begin{tabular}{c|c}
       $l$  & $f^{\mu\nu}_{\vec{l}L}$ \\
       \hline
        0 &  $-\dfrac{A^{\mu\nu}\,(1-i)}{\sqrt{\pi m_d}}\,\dfrac{\exp(-m_d\,(v-u))}{v-u}$\\
        1 &  $-\dfrac{A^{\mu\nu}\,(1+i)}{\sqrt{\pi m_d^3}}\,\dfrac{\exp(-m_d\,(v-u))}{(v-u)^3}\,(m_d\,(v-u)-1)$\\
        2 & $\dfrac{A^{\mu\nu}\,(1-i)}{\sqrt{\pi m_d}^5}\,\dfrac{\exp(-m_d\,(v-u))}{(v-u)^5}\,(3+3m_d\,(v-u)+m_d^2\,(v-u)^2)$\\
        3 & $\dfrac{A^{\mu\nu}\,(1+i)}{\sqrt{\pi m_d^7}}\,\dfrac{\exp(-m_d\,(v-u))}{(v-u)^7}\,(15+15\,m_d\,(v-u)+6\,m_d^2\,(v-u)^2+m_d^3\,(v-u)^3)$\\
        4 & $\dfrac{A^{\mu\nu}\,(1+i)}{\sqrt{\pi m_d^9}}\,\dfrac{\exp(-m_d\,(v-u))}{(v-u)^9}\,(105+105\,m_d\,(v-u)+45\,m_d^2\,(v-u)^2+10\,m_d^3\,(v-u)^3+m_d^4\,(v-u)^4)$
    \end{tabular}
    \caption{{The expression of the function $f^{\mu\nu}_{\vec{l}L}$ for some values of $l$.}}
    \label{tab:expr}
\end{table}

{From now on, the discussion follows closely that of \cite{art1}. Since we assume that there is no mixing between the extra dimensions and the 4d spacetime, we find that $B^{\alpha\beta}_{\vec{l}}$ can be expanded as follows:}
\begin{align}
B_{\vec{l}}^{00} &=\sum_{l\geq 0} a_{\vec{l}L}\, n_L\\\nonumber
B_{\vec{l}}^{0i} &= \sum_{l \geq 0} b_{\vec{l}L} n_L+\\
&+\sum_{l\geq 1} c_{\vec{l}iM-1}n_{M-1} + \epsilon_{iab}n_{aM-1}d_{\vec{l}bM-1}\\\nonumber
B_{\vec{l}}^{ij} &= \sum_{l\geq 0} \Big[ n_{ijM-2}e_{\vec{l}M-2}N_L + \delta_{ij} f_{\vec{l}L} n_{M}N_L\Big]+\\\nonumber
&+\sum_{l\geq 1} \Big[ n_{iM-1}g_{\vec{l}|j)M}+\epsilon_{ab(i}n_{j)M-1}h_{\vec{l}bM-1} \Big]+\\
&+\sum_{l\geq 2} \Big[ i_{\vec{l}ijM-2}n_{M-2}N_L + n_{aM-2}\epsilon_{\vec{l}ab(i}j_{j)bM-2} \Big]
\end{align}
while for the components $B^{AB}_{\vec{l}}$, we have
\begin{equation}\label{eq:AB}
\begin{split}
B_{\vec{l}}^{AB} &= \sum_{l} \Big[ e_{\vec{l}}N_{AB\hat{L}} + \delta_{AB}f_{\vec{l}}N_L \Big] +\\
&+\sum_{l}\Big[\sum_{i=3}^{d+1} N^i_{(A}c_{B)\vec{l}}N_{\vec{l}-1} + N^3_a\dots N^{d+1}_b\epsilon_{a\dots bc (A}N_{B)}d_{c\,\vec{l}-1}N_{\vec{l}-1} \Big]+\\
&+\sum_{l}\Big[N_{\vec{l}-2} a_{AB\vec{l}-2} + N^3_a\dots N^{d+1}_b\epsilon_{a\dots bc (A}b_{B)c\,\vec{l}-2}N_{\vec{l}-2} \Big].
\end{split}
\end{equation}

We now implement the condition $\partial_\mu h^{\mu\nu}=0$, which, with our assumptions, reduces to:
\begin{align}
&\partial_0 h^{00} + \partial_i h^{0i}=0\\
&\partial_0 h^{0i} + \partial_j h^{ij}=0\\
&\partial_B h^{AB}=0
\end{align}
If we define
\begin{align}
\mathcal{M}_{\vec{l}L} &=M_{\vec{l}L}\, \Bigg[ {A^{\mu\nu}\,(v-u)^{-1/2-l} J_{1/2+l}[i\,m_d\,(v-u)] -i\,A^{\mu\nu}\, (v-u)^{-1/2-l} Y_{1/2+l}[i\,m_d\,(v-u)]}\,{(v-u)^l} \Bigg]\\
\mathcal{S}_{\vec{l}L} &=S_{\vec{l}L}\, \Bigg[ {A^{\mu\nu}\,(v-u)^{-1/2-l} J_{1/2+l}[i\,m_d\,(v-u)] -i\,A^{\mu\nu}\, (v-u)^{-1/2-l} Y_{1/2+l}[i\,m_d\,(v-u)]}\,{(v-u)^l} \Bigg]\,
\end{align}
and introduce the following quantities {({see Section 2 of \cite{art1} for the details of the derivation}):}
\begin{align}
M_{\vec{l}L} &= a_{\vec{l}L}+2\,m_d\,b_{\vec{l},L+1}+m_d\,e_{\vec{l},L+1}+f_{\vec{l}L}\\
S_{\vec{l}L} &= -d_{\vec{l}L}-\dfrac{1}{2}\,m_d\,h_{\vec{l},L+1}\\
W_{\vec{l}L} &= b_{\vec{l}L}+\dfrac{1}{2}\,m_d\,e_{\vec{l},L+1}\\
X_{\vec{l}L} &= \dfrac{1}{2} e_{\vec{l}L}\\
Y_{\vec{l}L} &= -m_d\,b_{\vec{l},L+1}-m_d\,e_{\vec{l},L+1}-f_{\vec{l}L}\\
Z_{\vec{l}L} &= \dfrac{1}{2} h_{\vec{l}L},
\end{align}
{which can be obtained by deriving the Bessel functions.}
We find, analogously {to equations (2.28) in  \cite{art1}}
\begin{align}
c_{\vec{l}L} &= -m_d\,\Big( M_{\vec{l},L+1}+Y_{\vec{l},L+1} \Big)\\
g_{\vec{l}L} &= 2 Y_{\vec{l}L}\\
i_{\vec{l}L} &= m_d\,M_{\vec{l},L+1}\\
j_{\vec{l}L} &= 2\, S_{\vec{l},L+1}.
\end{align}
There are no similar relations for the part $h^{AB}$, which is unconstrained. From the above, we find, similarly to {equation (2.32) of \cite{art1}}:
\begin{align}
h^{00} =& -\dfrac{4}{c^2} \, \sum_{l\geq 0}\sum_{l_d\geq 1} \dfrac{(-1)^l}{l!} \mathcal{M}_{\vec{l},M}N_{\vec{l}}n_{M}\\\nonumber
h^{i0} =& +\dfrac{4}{c^3} \, \sum_{l\geq 1}\sum_{l_d\geq 1} \dfrac{(-1)^l}{l!}m_d\, \mathcal{M}_{\vec{l},iM-1} N_{\vec{l}}n_{M-1}+\dfrac{4}{c^2} \, \sum_{l\geq 1}\sum_{l_d\geq 1} \dfrac{(-1)^l\,l}{(l+1)!} \epsilon_{iab}\mathcal{S}_{\vec{l},bM-1}n_{aM-1}N_{\vec{l}}\\\nonumber
h^{ij} =& -\dfrac{4}{c^4} \sum_{l\geq 2}\sum_{l_d\geq 1} \dfrac{(-1)^l}{l!}  m_d^2\, \mathcal{M}_{\vec{l},M-2}N_{\vec{l}}n_{M-2}-\dfrac{8}{c^4} \sum_{l\geq 2}\sum_{l_d\geq 1} \dfrac{(-1)^l\,l}{(l+1)!} ,m_d\, \epsilon_{ab(i}\mathcal{S}_{\vec{l},bM-2}N_{\vec{l}}n_{aM-2}
\end{align}
plus gauge terms, see \cite{LRR1,art1}. The $l_d=m_d=0$ terms have the usual expression and a similar expansion as the one given above, so one might rewrite the above formulae with $l_d\geq 0$ in the inner sum, keeping in mind that the contributions for $l_d\geq 1$ contain a decaying exponential. Similarly, the $h^{AB}$ part of the perturbation tensor can be decomposed as in \eqref{eq:AB} and is valid for both $l_d=0$ and $l_d\geq 1$.

\section{General expression of the multipole expansion}
\label{sec:non-homog}

We need to solve the non-homogeneous wave equation:
\begin{equation}\label{eq:general}
    \Box u=\left( \dfrac{1}{c^2}\partial^2_t-\Delta-\Delta_d \right)\,u=S
\end{equation}
where $S$ is the source term. We look for a Green's function $\hat{G}$. We proceed as above, first expanding in terms of the higher dimensional spherical harmonics:
\begin{equation}
    \left( \dfrac{1}{c^2}\partial^2_t-\Delta+m_d^2 \right)\,\hat{G}_{\vec{l}}=4\pi\,\delta^4.
\end{equation}
The solution is:
\begin{equation}\label{eq:greend}
\begin{split}
    \hat{G}_{\vec{l}}(x,x^\prime)&=-\dfrac{\theta(t-|\vec{x}-\vec{x}^\prime|/c)}{2\pi}\,\delta(t^2-|\vec{x}-\vec{x}^\prime|^2/c^2)+\\
    &+\theta(t-|\vec{x}-\vec{x}^\prime|/c)\theta(t^2-|\vec{x}-\vec{x}^\prime|^2/c^2)\,m_d \dfrac{J_1(m_d\,\sqrt{t^2-|\vec{x}-\vec{x}^\prime|^2/c^2})}{4\pi\,\sqrt{t^2-|\vec{x}-\vec{x}^\prime|^2/c^2}}.
    \end{split}
\end{equation}
Recall the Dirac delta property:
\begin{equation}
\delta(c^2t^2-r^2) = \dfrac{1}{2r}\Big[ \delta(ct+r)+\delta(ct-r) \Big]
\end{equation}
The Bessel function has the following asymptotic expansions:
\begin{subequations}\label{eq:expansion}
\begin{equation}
    \dfrac{m_d^2\,r}{2},\qquad m_d\rightarrow0\, (R\rightarrow\infty),\\
\end{equation}
\begin{equation}    
    \dfrac{-8m_d^2\,r\,\cos\left( \pi/4+m_d\,r \right)+3m_d\,\sin\left( \pi/4+m_d\,r \right)}{\sqrt{32\pi}\,\left( m_d\,r \right)^{3/2}}, \qquad m_d\rightarrow\infty\,(R\rightarrow0)
\end{equation}
\end{subequations}

With this, equation \eqref{eq:general} has the following solution\footnote{We have to expand also the source term in the spherical harmonics.}
\begin{equation}
    u(t,\vec{r})=\int d^4x^\prime \hat{G}_{\vec{l}}(x,x^\prime)\,S_{\vec{l}}(x^\prime).
\end{equation}
There are two parts in the above expression, the one depending on $|\vec{r}-\vec{r}^\prime|^{-1}$ can be treated as usual {(see equations (6.3) and (6.4) in \cite{art1}, and see also \cite{LRR1,match1})}, thus obtaining:
\begin{subequations}
\begin{equation}
S_{\vec{l}}=S_{\vec{l},M}n_{M}
\end{equation}
\begin{equation}\label{eq:questa}
R_{\vec{l},M}(r,s) = r^l \int_0^r dx \dfrac{(r-x)^l}{l!} \left( \dfrac{2}{x} \right)^{l-1}\, S_{\vec{l},M}(x,s)
\end{equation}
\begin{equation}
\int d^3r^\prime \dfrac{S_{\vec{l},M}(t,\vec{r}^\prime)}{|\vec{r}-\vec{r}^\prime|} = \int_{-\infty}^{t-r} ds \partial_M \left\{ \dfrac{R_{\vec{l},M}(\frac{1}{2}(t-s-r),s)-R_{\vec{l},M}(\frac{1}{2}(t-s+r),s)}{r} \right\}
\end{equation}
\end{subequations}

The second part, depending on the Bessel function can be treated similarly introducing another function $R^\prime$:
\begin{subequations}
\begin{equation}\label{eq:questa2}
R_{\vec{l},M}^\prime(r,s) = r^l \int_0^r dx \dfrac{(r-x)^l}{l!} \left( \dfrac{2}{x} \right)^{l-1}\, \dfrac{J_1(m_d\,\sqrt{s(s+2x)})}{\sqrt{s(s+2x)}}S_{\vec{l},M}(x,s),
\end{equation}
\begin{equation}
\begin{split}
&\int d^3r^\prime \dfrac{J_1(m_d\,\sqrt{c^2t^2-r^2-(r^\prime)^2})}{\sqrt{c^2t^2-r^2-(r^\prime)^2}} S_{\vec{l},M}(t,\vec{r}^\prime) =\\
=& \int_{-\infty}^{t-r} ds \partial_M \left\{ \dfrac{R_{\vec{l},M}^\prime(\frac{1}{2}(t-s-r),s)-R_{\vec{l},M}^\prime(\frac{1}{2}(t-s+r),s)}{r} \right\}.
\end{split}
\end{equation}
\end{subequations}
Following \cite{art1}, in equations \eqref{eq:questa} and \eqref{eq:questa2}, the lower limit of the integration can be set to a generic $a$ which could be or not equal to zero, moreover, one can write:
\begin{equation}
\begin{split}
u_{\vec{l},M} &= \partial_{M}\left\{ \dfrac{1}{r}\int_{-\infty}^{t-r} ds \Big[ R_{\vec{l},M}\left(\frac{1}{2}(t-s-r),s\right) + R_{\vec{l},M}^\prime\left(\frac{1}{2}(t-s-r),s\right) \Big] \right\} +\\
&- \int_{-\infty}^{t-r} ds \partial_M \left\{ \dfrac{R_{\vec{l},M}(\frac{1}{2}(t-s+r),s) + R^\prime_{\vec{l},M}(\frac{1}{2}(t-s+r),s)}{r} \right\}
\end{split}
\end{equation}
If the source has compact support $r\leq r_0$ the second line of the above expression is zero and in the first, one can operate a change of variable and rewrite the integral as {in equation (140) in \cite{LRR1}}:
\begin{equation}
\begin{split}
&\int_{-\infty}^{t-r} ds \Big[ R_{\vec{l},M}\left(\frac{1}{2}(t-s-r),s\right) + R_{\vec{l},M}^\prime\left(\frac{1}{2}(t-s-r),s\right) \Big]=\\
=&\int d^3r \, \left\{ r_M\, \int_{-1}^{+1} dz \, \Big[ \delta_l(z) S_{\vec{l}}(\vec{r},u+z \, r/c) \Big] \right\}+\\
+&\int d^3r \, \left\{ r_M\, \int_{-1}^{+1} dz \, \left[ \delta_l(z) \left( \dfrac{J_1(m_d\,\sqrt{(u+zr/c)(u+zr/c+2r/c)})}{\sqrt{(u+zr/c)(cu+zr/c+2r/c)}} S_{\vec{l}}(\vec{r},u+z \, r/c) \right)\right] \right\}
\end{split}
\end{equation}
with
\begin{equation}
\delta_l(z) = \dfrac{(2l+1)!!}{2^{l+1}l!} \, (1-z^2)^l
\end{equation}

From this, following \cite{LRR1} and references therein, one can rewrite the multipole expansion of the perturbation as follows:
\begin{equation}
\mathcal{M}(h^{\mu\nu}_{\vec{l}}) = \mathcal{FP}_{B=0} \Box_d^{-1} \Big[ r^{B} \mathcal{M}(\Lambda^{\mu\nu}_{\vec{l}}) \Big] - \dfrac{4G}{c^4} \sum_l \left\{ \dfrac{(-1)^l}{l!} \, \partial_M \,\left[ \dfrac{1}{r} \mathcal{H}^{\mu\nu}_{\vec{l},M}(u) \right] \right\} + \text{Homogeneous}
\end{equation}
with
\begin{equation}\label{eq:multipole}
\begin{split}
\mathcal{H}^{\mu\nu}_{\vec{l},M}(u) &= \mathcal{FP}_{B=0} \int d^3r\, r^B\,r_M \, \Big[ \int_{-1}^1 \delta_l(z) \overline{\tau}^{\mu\nu}(\vec{r},u+zr/c) \Big]+\\
+&\mathcal{FP}_{B=0} \int d^3r \, \left\{ r_M\, \int_{-1}^{+1} dz \, \left[ \delta_l(z) \left( \dfrac{J_1(m_d\,\sqrt{(u+zr/c)(u+zr/c+2r/c)})}{\sqrt{(u+zr/c)(cu+zr/c+2r/c)}} \overline{\tau}^{\mu\nu}_{\vec{l}}(\vec{r},u+z \, r/c) \right)\right] \right\}
\end{split}
\end{equation}
where $\overline{\tau}^{\mu\nu}$ is the PN expansion of
\begin{equation}
|g|\, T^{\mu\nu} + \dfrac{c^4}{16\pi G} \Lambda^{\mu\nu}
\end{equation}
and where $\Box_d^{-1}$ is the Green function \eqref{eq:greend}. The \emph{Homogeneous} part is the homogeneous solution reported in section \ref{sec:homogeneous}.

Generalizing the expression reported in {equation (2.26a) in \cite{thorne}}, one has
\begin{equation}
S_{\vec{l}} = \dfrac{(2l_d+1)!!}{A_d\, l_d!} \int d\Omega^d\, N_{\vec{l}} S
\end{equation}
where $$A_d=\dfrac{2\pi^{d/2}}{\Gamma(d/2)}$$ is the area of $S^d$, therefore the final expression for the multipole is:
\begin{equation}
\mathcal{M}(h^{\mu\nu})= \sum_{l_d} \dfrac{(2l_d+1)!!}{A_d\, l_d!}\, \mathcal{M}(h^{\mu\nu}_{\hat{L}})
\end{equation}

Finally, the PN expansion of the inner integrals in the two lines of equation \eqref{eq:multipole} is:
\begin{align}\nonumber
\text{First line} &= \sum_{p=0}^{\infty} \dfrac{(2l+1)!!}{(2p)!!(2l+2p+1)!!} \, \dfrac{r^{2p}}{c^2} \, \dfrac{\partial^{2p}}{\partial u^{2p}} \overline{\tau}=\\
&= \overline{\tau} + \dfrac{1}{(2l+3)} \, \dfrac{r^2}{c^2}\, \dfrac{\partial^2 \overline{\tau}}{\partial u^2} + \dfrac{1}{8(2l+3)(2l+5)}\, \dfrac{r^4}{c^4} \, \dfrac{\partial^4 \overline{\tau}}{\partial u^4}+ \dots\\
\text{Second line} &= \sum_{p=0}^{\infty} \dfrac{(2l+1)!!}{(2p)!!(2l+2p+1)!!} \, \dfrac{r^{2p}}{c^2} \, \dfrac{\partial^{2p}}{\partial u^{2p}} \left(\dfrac{J_1(m_d\,\sqrt{(u+zr/c)(u+zr/c+2r/c)})}{\sqrt{(u+zr/c)(cu+zr/c+2r/c)}}\overline{\tau}\right)
\end{align}
The last is a complicated expression involving $J_n$ for $n=\{0,1,2,3\}$ for $p=\{0,1,2\}$. The above formulae are simplified if one considers the asymptotic expansions \eqref{eq:expansion}.

\section{Green's function of the Laplacian}
\label{sec:green}

As a final step before calculating the metric expansion, we need the Green's function of the Laplacian
\begin{equation}\label{eq:init}
    (\Delta+\Delta_d)\,G=4\pi\,\delta^{4+d}
\end{equation}
We proceed as above, and find:
\begin{equation}
    (\Delta-m_d^2)\,G_{\vec{l}}=4\pi\,\delta^{4}
\end{equation}
This has the following solution for $m_d\neq 0$
\begin{equation}
    G_{\vec{l}}(x,x^\prime)=\dfrac{1}{4\pi}\,\dfrac{\exp\left(-m_d\,|\vec{x}-\vec{x}^\prime|\right)}{|\vec{x}-\vec{x}^\prime|}.
\end{equation}
The case $m_d=0$ reduces to the usual one and we have the usual expansion
\begin{equation}
    G_{\vec{l}}(x,x^\prime)=\dfrac{1}{4\pi}\,\dfrac{1}{|\vec{x}-\vec{x}^\prime|}.
\end{equation}

{The full Green's function is given by}
\begin{equation}\label{eq:green_lap}
    G(x,x^\prime)=\dfrac{1}{4\pi}\,\sum_{l_d}\left[\dfrac{\exp\left(-m_d\,|\vec{x}-\vec{x}^\prime|\right)}{|\vec{x}-\vec{x}^\prime|}\right] N_{\vec{l}}.
\end{equation}

\section{The metric up to 1PN order}
\label{sec:met}

We assume that the matter is confined on the 4D spacetime, thus $T^{AB}=0$. With this assumption, up to 1PN order, we find the following expansion of the metric tensor:
\begin{align}
g_{00} &= -1+\dfrac{2}{c^2} V - \dfrac{1}{c^4}\Big(2 V^2+ V_{(2)}\Big)+ O(6)\\
g_{0i} &= -\dfrac{4}{c^3} V_i+ O(5)\\
g_{ij} &= \delta_{ij} \Bigg[ 1 + \dfrac{2}{(d+1)\,c^2} V  \Bigg]+ O(4)\\
g_{AB} &= \tilde{g}_{AB} + O(4)
\end{align}
The expression of $V_{(2)}$ is given by:
\begin{equation}
\begin{split}
    V_{(2)}&=-4\,l_d(l_d+d-1)\,V^2+\dfrac{d(d-7)}{d-2}\,\partial_iV\partial_iV+\dfrac{12-12d+d^2}{d-2}\,\partial_AV\partial_AV
\end{split}
\end{equation}
while all the other potentials have the usual expansions; {see Sections 7.5 and 3.2.6 in \cite{LRR1,LRR2}, respectively, }and references therein. We notice that at this order there is no correction to the extra dimensions metric; this is due to our assumption that the matter is confined to the 4D spacetime. Corrections will appear at higher PN orders.

\subsection{Quadratic terms}
In the calculation of the potentials, quadratic terms of the type $\partial V\partial V$ will appear. We need to distinguish two cases: $m_d=0$ and $m_d\neq 0$. In the first case, we need to solve equations of the type:
\begin{align}
    \Delta f&=\left(\partial_i^n\dfrac{1}{r_1}\right)\left(\partial_i^m\dfrac{1}{r_1}\right),\\
    \Delta g&=\left(\partial_i^n\dfrac{1}{r_1}\right)\left(\partial_i^m\dfrac{1}{r_2}\right).
\end{align}
These equations have been solved and the result is reported, for example, in \cite{art2}. The second case is more complicated. We need to solve equations of the type:
\begin{align}
    (\Delta -m_d^2)F&=\exp\left(-m_d^\prime\,r_1\right)\dfrac{1}{r^n_1}\exp\left(-m_d^{\prime\prime}\,r_1\right),\\
    (\Delta -m_d^2)G&=\dfrac{\exp\left(-m_d^\prime\,r_1\right)}{r_1^m}\,\dfrac{\exp\left(-m_d^{\prime\prime}\,r_2\right)}{r_2^n}.
\end{align}
We were able to find solutions to these equations and their expressions are as follows:
\begin{equation}
\begin{split}
    F\left(n\right)&=\dfrac{1}{n\,m_d}\,\exp\left( \dfrac{(m_d+m_d^\prime+m_d^{\prime\prime})\,r_1^n}{n}\right)\, \Bigg[ \exp\left( -\dfrac{2m_d\,r_1^n}{n} \right)\, \text{Ei}\left( \dfrac{(m_d-m_d^\prime+m_d^{\prime\prime})\,r_1^n}{n} \right)+\\
    &-\text{Ei}\left( -\dfrac{(m_d+m_d^\prime+m_d^{\prime\prime})\,r_1^n}{n} \right) \Bigg]
\end{split}
\end{equation}
\begin{equation}
    \begin{split}
        G\left( m;n \right)&=\dfrac{1}{m_d\,(m\,r_1^n-n\,r_2^m)}\,\exp\left( -\dfrac{(m_d^\prime+m_d^{\prime\prime})\,(m\,r_1^n+n\,r_2^n)}{mn} \right)\,\Bigg\{\exp\left( \dfrac{(m_d^\prime+m_d^{\prime\prime})\,r_1^n}{n} \right)\times\\
        &\times\left( \exp\left( \dfrac{m_d^\prime\,r_2^m}{m} \right)\,\text{Ei}\left( \dfrac{(m_d-m_d^\prime+m_d^{\prime\prime})\,r_2^m}{m} \right)-\exp\left( -\dfrac{(2m_d+m_d^\prime)\,r_2^m}{m} \right)\, \text{Ei}\left( -\dfrac{(m_d+m_d^\prime-m_d^{\prime\prime})\,r_2^m}{m} \right) \right)+\\
        &+\exp\left( \dfrac{(m_d+m_d^{\prime\prime})\,r_2^m}{m} \right)\, \Bigg[ \exp\left( \dfrac{(2m_d+m_d^\prime)\,r_1^n}{n}\right)  \,\Bigg( \text{Chi}\left( -\dfrac{(m_d+m_d^\prime-m_d^{\prime\prime})\,r_1^n}{n} \right)-\text{Shi}\left( \dfrac{(m_d+m_d^\prime-m_d^{\prime\prime})\,r_1^n}{n} \right) \Bigg)+\\
        &-\exp\left( \dfrac{m_d^\prime\,r_1^n}{n} \right)\,\Bigg( \text{Chi}\left( \dfrac{(m_d-m_d^\prime+m_d^{\prime\prime})\,r_1^n}{n} \right)+\text{Shi}\left( \dfrac{(m_d-m_d^\prime+m_d^{\prime\prime})\,r_1^n}{n} \right) \Bigg)\Bigg]\Bigg\}
    \end{split}
\end{equation}
The derivation is given in the appendix. In the above expressions Shi, Chi and Ei are, respectively, the sinh integral, the cosh integral, and the exponential integral. We shall also need the above expressions evaluated at the position of particle 1 (or 2): these expressions are:
\begin{equation}
\begin{split}
    F_1(n)&=\dfrac{1}{n\,m_d}\,\exp\left( \dfrac{(m_d+m_d^\prime+m_d^{\prime\prime})\,r^n}{n}\right)\, \Bigg[ \exp\left( -\dfrac{2m_d\,r^n}{n} \right)\, \text{Ei}\left( \dfrac{(m_d-m_d^\prime+m_d^{\prime\prime})\,r^n}{n} \right)+\\
    &-\text{Ei}\left( -\dfrac{(m_d+m_d^\prime+m_d^{\prime\prime})\,r^n}{n} \right) \Bigg]
\end{split}
\end{equation}
\begin{equation}
    \begin{split}
        G_1(n,m)&=\dfrac{1}{m_d\,n\,r^m}\,\exp\left( -\dfrac{(m_d+m_d^\prime+m_d^{\prime\prime})\,r^m}{m} \right)\,\Bigg[ \exp\left( \dfrac{m_d^{\prime\prime}\,r^m}{m} \right)\,\text{Ei}\left( \dfrac{(m_d+m_d^\prime+m_d^{\prime\prime})\,r^m}{m} \right)+\\
        &-\exp\left( \dfrac{(2m_d+m_d^{\prime\prime})\,r^m}{m} \right)\,\text{Ei}\left( -\dfrac{(m_d-m_d^\prime+m_d^{\prime\prime})\,r^m}{m} \right)+\\
        &+\exp\left( \dfrac{(m_d+m_d^\prime)\,r^m}{m} \right)\,\ln\left( \dfrac{m_d-m_d^\prime+m_d^{\prime\prime}}{m_d+m_d^\prime-m_d^{\prime\prime}} \right) \Bigg]
    \end{split}
\end{equation}

\subsection{The potentials}
With the above functions, using the usual procedure described in \cite{art2}, we find the following expression of the potentials:
\begin{equation}
\begin{split}
    V&=\int d^d\theta \sqrt{\tilde{g}_d}\,\dfrac{Gm_1}{r_1}\,\sum_{l_d}\exp\left( -m_d\,r_1 \right)\,Y_{l_1,\dots,l_d}(\vec{\theta})+\\
    &+\dfrac{1}{c^2}\dfrac{Gm_1}{r_1}\,\Bigg\{ \int d\theta\,\sqrt{\tilde{g_d}}\,\sum_{l_d}\exp\left(-m_d\,r_1\right)\,\left[ \dfrac{3v_1^2}{2}-\dfrac{Gm_2}{r}\,\int d^d\theta^\prime\,\sqrt{\tilde{g}_d}\,\sum_{l_d^\prime}\exp\left(-m_d^\prime\,r\right)\,Y_{l_1,\dots,l_d}(\vec{\theta}^\prime) \right]\,Y_{l_1,\dots,l_d}(\vec{\theta})+\\
&+\dfrac{Gm_1}{2} \left[ -a_1^i\partial_ir_1+v_1^iv_1^j\partial_{ij}r_1 \right] \Bigg\}
\end{split}
\end{equation}
\begin{equation}
    V_i=\int d^d\theta \sqrt{\tilde{g}_d}\,\dfrac{Gm_1\,v_1^i}{r_1}\,\sum_{l_d}\exp\left( -m_d\,r_1 \right)\,Y_{l_1,\dots,l_d}(\vec{\theta})
\end{equation}
\begin{equation}
    \begin{split}
V_{(2)}&=-4\sum_{l_d}\,l_d(l_d+d-1)\,\Bigg[ \dfrac{G^2m_1^2}{8}\,\ln(r_1)+G^2m_1m_2\,{}_ig_{match\,i} \Bigg]+\\
&-4\int d^d\theta \sqrt{\tilde{g}_d} \int d\theta^\prime \sqrt{\tilde{g}_d} \sum_{l_d,l_d^\prime,l_d^{\prime\prime}}\,l_d(l_d+d-1)\,\Big[ G^2m_1^2\,F(2)+G^2m_1m_2\,G(1,1) \Big]\,Y_{l_1,\dots,l_d}(\vec{\theta})\,Y_{l_1,\dots,l_d}(\vec{\theta}^\prime)+\\
&+\dfrac{d(d-7)}{d-2}\,\left( \dfrac{G^2m_1^2}{8}\,\left( \Delta\ln(r_1)+\dfrac{3}{r_1^2} \right)+G^2m_1m_2\,{}_ig_{match\,i} \right)+\\
&+\dfrac{12-12d+d^2}{d-2}\,\int d^d\theta \sqrt{\tilde{g}_d} \int d\theta^\prime \sqrt{\tilde{g}_d}\sum_{l_d,l_d^\prime,l_d^{\prime\prime}}\Big( G^2m_1^2\,F(2)+G^2m_1m_2\,G(1,1) \Big)\,\partial_AY_{l_1,\dots,l_d}(\vec{\theta})\,\partial_AY_{l_1,\dots,l_d}(\vec{\theta}^\prime)+\\
&+\dfrac{d(d-7)}{d-2}\,\int d^d\theta \sqrt{\tilde{g}_d} \int d\theta^\prime \sqrt{\tilde{g}_d} \sum_{l_d,l_d^\prime,l_d^{\prime\prime}}\,\Bigg[ G^2m_1^2\,(\,F(4)+(m_d^\prime+m_d^{\prime\prime})\,F(3)+m_d^\prime m_d^{\prime\prime}\,F(2))+\\
&+G^2m_1m_2\,\left(G(3,3)+m_d^\prime\,G(2,3)+m_d^{\prime\prime}\,G(3,2)+m_d^\prime m_d^{\prime\prime}\,G(2,2)\right)\,r_1\,r_2\,(n_1n_2) \Bigg]\,Y_{l_1,\dots,l_d}(\vec{\theta})\,Y_{l_1,\dots,l_d}(\vec{\theta}^\prime)
    \end{split}
\end{equation}
{The function $g_{match}$ is defined in Section IV.C.3 of reference \cite{art2}. In order to derive the above expressions, one should remember that the inverse Laplacian is given by equation \eqref{eq:green_lap} which contains $m_d=\sqrt{l_d(l_d+d-1)}/R$. Considering for definiteness the expression for $V$, we have that at second order another $m_d$ appears, coming from the leading order of the expression (see the general expression of the potential $V$ in equations (4.4)--(4.6) of \cite{art2}). This, in principle, is different from the $m_d$ of the inverse Laplacian: we indicate this fact with a prime. We have the same situation in $V_{(2)}$, but now three different $m_d$ appear, thus, we use a prime and double prime notation in order to differentiate them. We need to sum over all these primed terms.} {Finally, in the expressions for the potentials reported above, sums appear over the eigenvalues of the higher order spherical harmonics $Y_{\vec{l}}(\vec{\theta})$ and integrals over the angular coordinates of the extra dimensions. These are angular integrals and do not require integrations by part, simplifying the treatment of the extra dimensions. This is the result of the expansion of the inverse Laplacian \eqref{eq:green_lap} into higher order spherical harmonics; this would not be true anymore had we chosen to solve the full equation \eqref{eq:init} without expansion.}

\section{Acceleration and equations of motion}

The acceleration of particle 1 is given by \cite{art1,LRR1,LRR2}
\begin{equation}
    a_1^i=F_1^i-\dfrac{d}{dt}\,\left( P_1^i -v_1^i \right)
\end{equation}
where {the force density and the linear momentum density relative to particle 1 are given, respectively, by}
\begin{align}
    F_1^i&=\dfrac{1}{2}\,\dfrac{[\partial_i g_{\mu\nu}]_1v_1^\mu v_1^\nu}{\sqrt{-[g_{\rho\sigma}]_1\,\dfrac{v_1^\rho v_1^\sigma}{c^2}}},\\
    P_1^i&=\dfrac{[g_{i\nu}]_1 v_1^\nu}{\sqrt{-[g_{\rho\sigma}]_1\,\dfrac{v_1^\rho v_1^\sigma}{c^2}}},
\end{align}
{where $v_1^i$ are the components of the particle 1 velocity.} Explicitly, we find:
\begin{align}
    F_1^i&=\left[\partial_iV\right]_1+\dfrac{1}{2c^2}\left( -2\left[V\right]_1\left[ \partial_iV \right]_1 + 3v_1^2\,\left[ \partial_iV \right]_1 -8 \left[ V_i \right]_1v_1^i-\left[ \partial_iV_{(2)} \right]_1 \right) +O(c)^{-4},\\
    P_1^i&=v_1^i+\dfrac{1}{2c^2}\,\left( 6\left[V\right]_1\,v_1^i+ v_1^2\,v_1^i - 8 \left[ V_i \right] _1\right)+O(c)^{-4}
\end{align}
At this order, the only difference with respect to the usual expression is the term $-\left[ \partial_iV_{(2)} \right]_1$, see {equation (3.35) in \cite{art1}}. At this point, we can calculate the (long) expression of the acceleration evaluated at the position of the particles and in the center of mass. {We report it in the appendix \ref{app:acc}.} {We now define the quantity (see \cite{LRR1} and references therein)}
\begin{equation}
    \gamma=\exp\left( -{m_d\,r} \right)\dfrac{Gm}{c^2r},
\end{equation}
{The $r$ appearing above is the binary separation distance and is different from the general radial distance used in Section \ref{sec:homogeneous}.} It is possible to invert the above expression to obtain $r$ as a function of $\gamma$ using the Lambert function, however, it is also possible to obtain the following series expansion:
\begin{equation}
    r=\dfrac{G m}{c^2\,\gamma}\,\left( 1-\dfrac{G m\,m_d}{c^2\gamma}+\dfrac{3}{2}\,\dfrac{G^2m^2\,m_d^2}{c^4\,\gamma^2}-\dfrac{8}{3}\,\dfrac{G^3m^3\,m_d^3}{c^6\gamma^3} +O(c)^{-8}\right),
\end{equation}
which reduces to the usual expression for $m_d=0$. If we consider circular orbits and substitute the series expansion for $r$ in the expression for the acceleration in the center of mass and expand the resulting expression up to the order $O(c^{-2})$, we arrive at the expression
\begin{equation}
    \vec{a}=-\Omega^2\,\vec{x}=-\dfrac{Gm}{r^3}\,\vec{x}\,\int d^d\theta \sum_{l_d} \sqrt{\tilde{g}_d}\Bigg[ -\dfrac{\gamma_p}{\gamma} + \left[1 - \gamma_p\,(1+4\nu) \right] +\gamma\,(\nu-3) \Bigg]\,Y_{l_1,\dots,l_d}(\vec{\theta}),
\end{equation}
where
\begin{equation}\label{eq:gp}
    \gamma_p=\dfrac{Gm\,m_d}{c^2},
\end{equation}
where $m$ is the total mass of the system. The above expression for the acceleration reduces to the one reported in \cite{LRR1} for $\gamma_p\rightarrow0$. We notice the presence of a term containing the inverse power of $\gamma$, corresponding to a term at -1PN order. The origin of this term can be traced back to the order zero term in the acceleration, i.e.
\begin{equation}
    \exp\left( -{m_d\,r} \right)\,\left[ 1+\,{m_d\,r} \right]\,\dfrac{G m}{r^2}\,n^i
\end{equation}
in particular from the second term in the square brackets. This last term appears because of the gradient of the potential $V$ applied to the exponential. This -1PN term will dominate the PN expansion of the waveform, unless it is very small. We think this is a major observational handle in the study of extra dimensions, since it strongly constrains their size and the number. 

The value of the term $\gamma_p=\sqrt{l_d(l_d+d-1)}\,Gm/{Rc^2}$ summed over $l_d$ and integrated over the extra dimensions is reported in table \ref{tab:valori} for different numbers of dimensions $d$. We leave the dependence over $R$ and the total mass $m$ indicated. We notice that it gets smaller and smaller the more the number of extra dimensions increases. 

\begin{table}[ht]
    \centering
    \begin{tabular}{c|c||c|c}
        d & $\gamma_p$ & d & $\gamma_p$ \\
        \hline
        3 & $3.8\,\times\,10^{-27}\,\dfrac{m}{R}$ & 6 & $9.6\,\times\,10^{-32}\,\dfrac{m}{R}$\T\B\\
        4 & $6.4\,\times\,10^{-28}\,\dfrac{m}{R}$ & 7 & $5.0\,\times\,10^{-35}\,\dfrac{m}{R}$\T\B\\
        5 & $2.1\,\times\,10^{-29}\,\dfrac{m}{R}$ & 8 & $2.5\,\times\,10^{-39}\,\dfrac{m}{R}$\T\B\\
    \end{tabular}
    \caption{Values of $\gamma_p$ as a function of $d$}
    \label{tab:valori}
\end{table}

\section{The energy flux of gravitational waves}
\label{sec:flux}

In order to calculate the flux, we follow \cite{flux,flux2}. {In order to calculate the multipoles we first expand all the integrands over the symmetric trace free basis $N_{\vec{l}}$ from equation \eqref{eq:N}, in this way, all the integrals are reduced to integrals over the three dimensional macroscopic space and no problematic integration by parts is necessary in the extra dimensions.} 

At the first PN order, we do not find any change in the multipole moments, but the flux is given by:
\begin{equation}
    \begin{split}
        \mathcal{F}&=\dfrac{32c^5\,\gamma^5\, \nu^2}{5G}\, \Big( \dfrac{a}{\gamma^2}+\dfrac{b}{\gamma}+c+d\,\gamma\Big)+O(\gamma)^7
    \end{split}
\end{equation}
where
\begin{align}
    a&=\sum_{l_d}\gamma_p^2\,N_{\vec{l}},\\
    b&=-\sum_{l_d}\left[\dfrac{-42\,\gamma_p+\gamma_p^2\,(43+207\,\nu)}{21}\right]\,N_{\vec{l}},\\
    c&=\dfrac{1764+\sum_{l_d}\,\left[ 168\,\gamma_p\,(41-144\,\nu)+\gamma_p^2\,(1933+21414\,\nu+55953\,\nu^2) \right]\,N_{\vec{l}}}{1764},\\
    d&=-\dfrac{2927+420\,\nu}{336}-\dfrac{\sum_{l_d}\,\left[ \gamma_p\,(5459+25458\,\nu-24957\,\nu^2)+\gamma_p^2\,(1+4\nu)(1+39\nu)(43+207\nu)\right]\,N_{\vec{l}}}{882},
\end{align}
which reduces to the usual expression reported in \cite{LRR1,LRR2,flux,flux2} in the limit ${m_d\rightarrow0}$, {which is equivalent to} ${\gamma_p\rightarrow0}$. The summations are over the $l_d$ that appears in the definition of $\gamma_p$ \eqref{eq:gp}. We notice the presence of two extra terms depending on the powers $\gamma^{-2}$ and $\gamma^{-1}$ which are not present in the usual expression and originate from the inverse power of $\gamma$ in the equations of motion.

\section{Comparison to previous works}
\label{sec:comp}

{\cite{previous} performs a treatment of a binary system in a 5d setting using a Kaluza-Klein (KK) reduction of the metric. The treatment performed here is somewhat different from that in \cite{previous} and their results are also different. First of all authors of \cite{previous} use a KK reduction because, they say, it is not possible to perform the usual treatment in higher dimensions because of the difficulties in dealing with higher dimensional integration by parts. In our treatment, we think we have solved this issue with our decomposition \eqref{eq:homog2}, in which all integrations by part in extra dimensions disappear.}

{Authors of \cite{previous} consider a more general setting then we do. Their metric, after the KK reduction looks like this}
\begin{equation}
    G_{MN}=\exp\left(-\dfrac{\varphi}{3}\right)\,\left( \begin{array}{cc}
        g_{\mu\nu}+\kappa\,A_\mu A_\nu & \sqrt{\kappa} \exp(\varphi)\,A_\mu \\
        \sqrt{\kappa}\,\exp(\varphi)\,A_\nu & \exp(\varphi)
    \end{array} \right).
\end{equation}
{Compared with our metric, we see that we have considered a special situation in which the Maxwell field is zero, $A_\mu=0$. This could be an explanation of why we find that all the extra dimensions effects disappear when considering the limit $l_d,\gamma_p\rightarrow0$, while in their case they find that the potentials are much different from the 4d case even when the extra dimensions effects are neglected (and therefore conclude that extra dimensions are ruled out by observations).}

{We can compare our Newtonian potential with that of \cite{previous}, their equation (3.10); however, we need to sum all the terms over all the values of $0\leq m_q \leq \infty$ before comparing our results with those of \cite{previous}. We also need to perform all the integrals over the higher dimensional spherical harmonics in the potentials before comparing them with their $V$. The non trivial task is the summation over $l_d$, since we were unable to obtain an analytic expression for the sum for generic number of dimensions. In the 5d case, the integration is trivial, since only the term $l_d=0$ survives and the integral is 1; remembering the definition}
\begin{equation}
    m_d=\dfrac{\sqrt{l_d(l_d+d-1)}}{R},
\end{equation}
{one is left with the sum (in fact, a harmonic series)}
\begin{equation}
    \sum_{l_d=0}^\infty\dfrac{\exp(-m_d\,r)\,G\,m}{r}=\sum_{l_d=0}^\infty\,\dfrac{Gm}{r}\,\exp\,\left(-l_d\,\dfrac{r}{R}\right)=\dfrac{Gm}{r}\,\left[ 1-\exp\left( -\dfrac{r}{R} \right) \right]^{-1}.
\end{equation}
{Since \cite{previous} fixes $G=4G^{(5)}/(3R)$, the first term above is similar to that of \cite{previous}, however, even considering a series expansion for small $R$, the second term is not: a factor 2 is missing and we were unable to find the cause of this discrepancy. As discussed above, we consider block-diagonal metrics, while \cite{previous} doesn't, so this might be the source of the missing factor 2, but we need to repeat our work using more general metrics to be sure.}

{To end the comparison with \cite{previous}, the authors find a correction to the mass of a point like particle due to the presence of the dilatonic field (compare to equation (4.9) of \cite{previous})}
\begin{equation}\label{eq:mass}
    m_{\text{effective}}=m\,\exp\left(-\dfrac{\varphi}{6}\right)
\end{equation}
{We also find this correction. In fact if, as we assume, there is no motion of the particle in the extra dimensions, we find that the action is given by}
\begin{equation}
    S=-m\,\int d\tau\,\sqrt{\dot{x}^{\mu}\dot{x}^{\nu}\,g_{\mu\nu}+\dot{x}^A\dot{x}^B\,g_{AB}},
\end{equation}
{where the second term disappears, by assumption; comparing the metrics in the two papers, we see that our $g_{\mu\nu}$ is the term $\exp(-\varphi/3)\,g_{\mu\nu}$ in \cite{previous}. Substituting above, we find again equation \eqref{eq:mass}.}

\section{Conclusion and discussion}
\label{sec:concl}
We have presented a calculation of the metric, the equations of motion and of the energy flux of gravitational waves emitted by a compact binary in the case in which the spacetime contains $d$ extra dimensions compactified to a hypersphere. We have solved the homogeneous and the non-homogeneous wave equation in $d+4$ dimensions, showing that there is an infinite sum of pseudo-massive modes which decay exponentially fast at infinity plus the usual term decaying as $1/r$ which radiates to infinity. Calculating the metric at 1PN order, we have found that it is modified by the presence of a new potential. The equations of motion are also modified with the appearance of a new term at -1PN order. This, in turn, implies a modification of the energy flux. By comparing these findings with observations, it could be possible to estimate the number of the extra dimensions and their size.

{One of the assumptions in this work is that the $4+d$-dimensional metric is block diagonal. If we look at the definition of $\Lambda^{\mu\nu}$ given in equation (175) of reference \cite{LRR1} and use mixed indices $\mu\nu=iA$, we see that at least up to $O(c^{-4})$ there is no mixing between the macroscopic and the extra dimensions. However, a mixing might appear at higher orders. This would invalidate some of the results reported in this work. This is matter for a future work.}

{There are several works that study GW in higher dimensional spacetime, see the account \cite{review}. Many works deal with primordial GW \cite{randal1,randal2,randal3,randal4}, because the effects of GW are increased by the higher energy of the plasma in the young Universe, while at lower energies, when the Universe is older and colder, GW in higher dimensional spacetime are essentially indistiguishable from those in 4d spacetime. Papers dealing with binary systems are, for example, \cite{ap4,obs1}; see also references therein. In general, they find that the attenuation of GW is larger in higher dimensional spacetime, and a measurement of this attenuation could be a handle on the study of extra dimensions. Moreover, electromagnetic and gravitational waves can propagate differently in higher dimensional spacetime, thus multi-messenger observations can be used to constrain them. A paper that tries to constrain observationally the size and number of extra dimensions is \cite{ap4}. The authors use the fact that GW might propagate in other dimensions, while other fields, specifically the electromagnetic field, cannot; this means that the trajectory in the 4d brane is different for the two fields. They focus on the event GW170817/GRB 170817A and find that $d\leq9$, or if one wants to be conservative $d\leq12$. They are also able to put constraints on the size of the dimensions, finding (in our notation) $R^2\leq3.65\times10^{-54}$ Mpc$^{2}$, i.e, $R\leq58$ $\mu$m, which they find to be compatible with torsion balance results reported in \cite{tors}.} {In this context, with our paper, we give a further handle in the study of higher dimensions with GW, i.e., the -1PN order term. This term is very powerful in constraining the extra dimensions, since if they are present but very small, this term would be much larger than the others; in particular, it would be much larger than the Newtonian term. This would certainly have already been noticed in any of the about 400 GW events we have observed up to now with the ground-based observatories \cite{cata0,cata1,cata2,cata3,cata4}; thus, one should conclude that either the extra dimensions do not exist at all (in our equations this would be the limit $l_d, m_d\rightarrow0$), or that they are larger than expected (but they cannot be too large, since there are upper limits on their size, see, for example, \cite{eotwash,tors}), or, finally, that there is a sufficiently large number of extra dimensions to compensate for their size (see Table \ref{tab:valori}).}

{We compared our results with those of \cite{previous}, finding that in the case of one extra dimension, the results are similar, but not identical; however, we were unable to find the cause of the discrepancies. The source might be the fact that we consider a block-diagonal metric, while \cite{previous} doesn't. We need to repeat our calculations including also the non diagonal terms.}

\appendix

\section{Derivation of the expressions of $F(n)$ and $G(m,n)$}
We first write
\begin{equation}
    |\vec{x}_1-\vec{x}^\prime|=|\vec{x}_1-\vec{x}+\vec{x}-\vec{x}^\prime|=|\vec{r}_1-\vec{r}^\prime|.
\end{equation}
Asymptotically, this can be written as follows:
\begin{equation}
    |\vec{x}_1-\vec{x}^\prime|\approx r_1-\vec{n}_1\cdot\vec{r}^\prime.
\end{equation}
We substitute this into the integral:
\begin{equation}
\begin{split}
    &\int d^3x^\prime\,\exp\left( -m_d\, r^\prime\right)\,\dfrac{1}{r^\prime}\, \exp\left(-(m_d^\prime+m_d^{\prime\prime}) \big({r_1-\vec{n}_1\cdot\vec{r}^\prime}\big)\right) \dfrac{1}{r_1^n}=\\
    =&\exp\left(-(m_d^\prime+m_d^{\prime\prime})\, {r_1}\right) , \int d^3x^\prime\exp\left( -m_d \,r^\prime\right)\,\dfrac{1}{r^\prime}\,\dfrac{1}{r_1^n-\vec{n}\cdot\vec{r}^\prime}\, \exp\left((m_d^\prime+m_d^{\prime\prime})\, {\vec{n}_1\cdot\vec{r}^\prime}\right).
    \end{split}
\end{equation}
Integrating with Mathematica, we can find the given formula for $F(n)$.

For the formula $G(m,n)$, we proceed analogously by expanding the exponentials. We arrive at
\begin{equation}
    \begin{split}
        \exp\left( -m_d^\prime\,r_1-m_d^{\prime\prime}\,r_2 \right)\int &dx^\prime\,\exp\left( -\sqrt{l_d(l_d+d-1)}\, \dfrac{r^\prime}{R}\right)\,\dfrac{1}{r^\prime}\, \dfrac{\exp\left( m_d^\prime\,\vec{n}_1\cdot\vec{r}^\prime+m_d^{\prime\prime}\,\vec{n}_2\cdot\vec{r}^\prime \right)}{(r_1^m-m\,\vec{n}_1\cdot\vec{r}^\prime)(r_2^n-n_2\,\vec{n}\cdot\vec{r}^\prime)}.
    \end{split}
\end{equation}
We now have two angles. We consider the reference frame with the origin in the center of mass of the system, thus $\vec{n}_1\cdot\vec{r}^\prime=-\vec{n}_2\cdot\vec{r}^\prime=\vec{n}\cdot\vec{r}^\prime$, and the above formula simplifies:
\begin{equation}
    \begin{split}
        \exp\left( -m_d^\prime\,r_1-m_d^{\prime\prime}\,r_2 \right)\int &dx^\prime\,\exp\left( -\sqrt{l_d(l_d+d-1)}\, \dfrac{r^\prime}{R}\right)\,\dfrac{1}{r^\prime}\, \dfrac{\exp\left( (m_d^\prime-m_d^{\prime\prime})\,\vec{n}\cdot\vec{r}^\prime\right)}{(r_1^m-m\,\vec{n}\cdot\vec{r}^\prime)(r_2^n+n\,\vec{n}\cdot\vec{r}^\prime)}.
    \end{split}
\end{equation}
This can be integrated as above, and the result is the given function $G(m,n).$

\section{The acceleration in the center of mass}
\label{app:acc}
{In this appendix, we report the long expression of the acceleration in the center of mass at 1PN. It is gven by:}
\begin{equation}
    a^i=-\exp\left( -\dfrac{m_d\,r}{R} \right)\,\dfrac{Gm}{r^2}\, \left[ \left(1 +\dfrac{m_d\,r}{R}+\dfrac{1}{c^2}\,\mathcal{A}\right)\,n^i+\dfrac{1}{c^2}\,\mathcal{B}\,v^i \right]
\end{equation}
{where}
\begin{equation}
    \begin{split}\nonumber
        \mathcal{A}&=\left[ -\dfrac{3\,\nu\,\dot{r}^2}{2} - \dfrac{2Gm}{r}\,\left(2+\nu+\dfrac{(R+m_dr)\,\nu}{R\, r}\right)+2+v^2\,\left(1+3\nu+\dfrac{3m_d\,r}{R}\right)-\dfrac{G m \,\ m_d}{R}\exp\left( -\dfrac{m_d\,r}{R} \right) \right]+\\\nonumber
        &+\dfrac{12-12d+d^2}{2(d-2)}\,\sum_{l_d,l_d^\prime,l_d^{\prime\prime}} \int d^d\theta_1 \sqrt{\tilde{g}_d} \int d^d\theta _2\sqrt{\tilde{g}_d} \Bigg\{\dfrac{Gm\,r^3}{m_d}\,\exp\left( \dfrac{(2m_d\,r+(m_d^\prime+m_d^{\prime\prime})\,r}{2} \right)\\
        &\Bigg[ (m_d-m_d^\prime-m_d^{\prime\prime})\,\text{Ei}\left( \dfrac{(m_d-m_d^\prime-m_d^{\prime\prime})^2\,r^2}{2} \right)+\\\nonumber
        &+\exp\left( {m_d\,r} \right)\,(m_d-m_d^\prime+m_d^{\prime\prime})\,\text{Ei}\left( -\dfrac{(m_d+m_d^\prime+m_d^{\prime\prime})^2\,r^2}{2} \right) \Bigg]\Bigg\}\,\partial_AY_{l_1,\dots,l_d}(\vec{\theta}_1)\,\partial_AY_{l_1,\dots,l_d}(\vec{\theta}_2)+\\\nonumber
        &-4m_d^2\,\sum_{l_d,l_d^\prime,l_d^{\prime\prime}} \int d^d\theta_1 \sqrt{\tilde{g}_d} \int d^d\theta _2\sqrt{\tilde{g}_d} \Bigg\{\dfrac{Gm\,r^3}{m_d}\,\exp\left( \dfrac{(2m_d\,r+(m_d^\prime+m_d^{\prime\prime})\,r}{2R} \right)\\
        &\Bigg[ (m_d-m_d^\prime-m_d^{\prime\prime})\,\text{Ei}\left( \dfrac{(m_d-m_d^\prime-m_d^{\prime\prime})^2\,r^2}{2} \right)+\\\nonumber
        &+\exp\left( {m_d\,r} \right)\,(m_d-m_d^\prime+m_d^{\prime\prime})\,\text{Ei}\left( -\dfrac{(m_d+m_d^\prime+m_d^{\prime\prime})^2\,r^2}{2} \right) \Bigg]\Bigg\}\,Y_{l_1,\dots,l_d}(\vec{\theta}_1)\,Y_{l_1,\dots,l_d}(\vec{\theta}_2)+\\\nonumber
        &+\dfrac{d(d-7)}{d-2}\sum_{l_d^\prime,l_d^\prime,l_d^{\prime\prime}} \int d^d\theta_1 \sqrt{\tilde{g}_d} \int d^d\theta _2\sqrt{\tilde{g}_d} \Bigg\{\,Gm\,r^3\,\Bigg[ 6\,\exp\left( \dfrac{(-m_d+m_d^\prime+m_d^{\prime\prime})^2\,r^2}{2} \right)\,m_d^\prime\,m_d^{\prime\prime}\,\\
        & (m_d-m_d^\prime+m_d^{\prime\prime})\,\text{Ei}\left( \dfrac{(-m_d+m_d^\prime+m_d^{\prime\prime})^2\,r^2}{2} \right)+\\\nonumber
        &+\exp\left( \dfrac{m_d^2\,r^2}{2} \right)\,(m_d+m_d^\prime+m_d^{\prime\prime})\,\text{Ei}\left( -\dfrac{(m_d+m_d^\prime+m_d^{\prime\prime})^2\,r^2}{2} \right) +\\\nonumber
        &+4\,\exp\left( \dfrac{(-m_d+m_d^\prime+m_d^{\prime\prime})^3\,r^3}{3} \right)\,(m_d^\prime+m_d^{\prime\prime})\,R\,r\, (m_d-m_d^\prime+m_d^{\prime\prime})\,\text{Ei}\left( \dfrac{(-m_d+m_d^\prime+m_d^{\prime\prime})^3\,r^3}{3} \right)+\\\nonumber
        &+\exp\left( \dfrac{2m_d^3\,r^3}{3} \right)\,(m_d+m_d^\prime+m_d^{\prime\prime})\,\text{Ei}\left( -\dfrac{(m_d+m_d^\prime+m_d^{\prime\prime})^3\,r^3}{3} \right)+\\\nonumber
        &+3\,\exp\left( \dfrac{(-m_d+m_d^\prime+m_d^{\prime\prime})^4\,r^4}{4} \right)\,R^2\,r^2\,\Bigg( (m_d-m_d^\prime+m_d^{\prime\prime})\,\text{Ei}\left( \dfrac{(-m_d+m_d^\prime+m_d^{\prime\prime})^4\,r^4}{4} \right)+\\\nonumber
        &+\exp\left( \dfrac{m_d^2\,r^2}{2} \right)\,(m_d+m_d^\prime+m_d^{\prime\prime})\,\text{Ei}\left( -\dfrac{(m_d+m_d^\prime+m_d^{\prime\prime})^4\,r^4}{4} \right)\Bigg)\Bigg]\Bigg\}\,Y_{l_1,\dots,l_d}(\vec{\theta}_1)\,Y_{l_1,\dots,l_d}(\vec{\theta}_2)
    \end{split}
\end{equation}

\begin{equation}
    \begin{split}
        \mathcal{B}&=-\dfrac{2\dot{r}}{R}\,\left( R+m_d\,r \right)\,(2-\nu)+O(c)^{-2}
    \end{split}
\end{equation}
where $\dot{r}=(nv)$ {and, as usual, $\nu$ is the symmetric mass ratio (see, for example equation (215) of \cite{LRR1})}. It can be checked, taking the limits for $l_d\rightarrow 0$ and $d\rightarrow0$ that this expression reduces to the usual one given, for example, in \cite{LRR1,LRR2}.

\bibliography{biblio}

\end{document}